\documentclass[a4paper]{article}
\usepackage{multicol}
\usepackage{amsmath}
\usepackage{mathrsfs}
\usepackage[utf8]{inputenc}
\usepackage{hyperref}
\usepackage{cite}
\usepackage{graphicx}
\usepackage[left=20mm, top=20mm, right=20mm, bottom=20mm, nohead, nofoot]{geometry}
\usepackage{caption}

\newcommand{\s}[1]{\hat \sigma_{#1}}
\newcommand{\p}[1]{\hat p_{#1}}
\newcommand{\vF}{v_\text{\tiny F}}
\newcommand{\Ds}[1]{\Delta_\text{s}{#1}}
\newcommand{\ew}{\varepsilon_\text{w}}
\newcommand{\com}{\mspace{5mu}\text{,}}
\newcommand{\pnt}{\mspace{5mu}\text{.}}
\newcommand{\fB}[1]{\Omega_\text B^{#1}}
\newcommand{\fpl}[1]{\omega_\text{pl}^{#1}}
\begin{document}
	\fontfamily{cmr}
	\selectfont
\onecolumn
\begin{center}
	\large{\textbf{Influence of the Constant Electric Field on the Mutual Rectification of the Electromagnetic Waves in Graphene Superlattice}}
\end{center}
\vspace{1ex}
\begin{center}
	S.V. Kryuchkov $^\text {a, b, }$\footnote{ svkruchkov@yandex.ru} and E.I. Kukhar’ $^\text {a, }$\footnote{ eikuhar@yandex.ru}
\end{center}
\vspace{1ex}
\begin{center}
	$^\text a$\textit{Volgograd State Socio-Pedagogical University, Physical Laboratory of Low-Dimensional Systems\footnote{ \url{http://edu.vspu.ru/physlablds}}, V.I. Lenin Avenue, 27, Volgograd 400066, Russia}
\vspace{2ex}
	
	$^\text b$\textit{Volgograd State Technical University, V.I. Lenin Avenue, 28, Volgograd 400005, Russia}
\end{center}

\begin{abstract}
In tight binding approximation the explicit form of the electron energy in graphene superlattice was derived.
The possibility of propagation of the cnoidal waves in graphene superlattice was discussed.
The direct current induced perpendicularly to the superlattice axis by cnoidal and sinusoidal electromagnetic waves under the presence of longitudinal constant electric field was calculated.
Such direct current was shown to change its direction when the intensity of longitudinal electric field changes its absolute value.
\\
\\
\\
\end{abstract}

\begin{multicols}{2}

\begin{center}\textbf{I. Introduction}\\\end{center}

The electronic properties of graphene-based structures in particular -- the features of electronic transport in graphene superlattices (GSL) -- are under the very intensive study presently \cite{1,2,3,4,5,6,7,8,9,10,11}.
The possibility of using of the superlattice (SL) as a working medium of generators and amplifiers of terahertz (THz) electromagnetic (EM) waves \cite{12} induces the investigations of electronic and optical features of carbon systems with the additional SL potential \cite{7,13,14,15}.
Dispersion low of graphene deposited on the periodic substrate (h-BN/SiO$_{2}$) is studied in \cite{9} where the energy of electron motion along the SL axis was shown to have the periodical dependence on the quasimomentum directed along SL axis.

In \cite{16,17,18,19,20,21,22,23} amplification of even harmonics of the polychromatic pump field in SL was shown to be accompanied by the appearance of a constant component of the field – so-called mutual rectification due to wave mixing \cite{24,25,26,27,28,29,30,31,32}.
Further such features of electron spectrum of GSL as periodicity and nonadditivity are shown to change the character of wave mixing effect in GSL and to be a cause of new peculiarities different from that of in bulk SL \cite{31,32}.

The expression of dispersion low in GSL obtained in \cite{9} gives the information about features of band structure of GSL.
However it can’t be used to calculate analytically the current induced in GSL by the external EM field because of the energy dependence on the quasimomentum is implicit in \cite{9}.
In \cite{33} the numerical analysis of dispersion low was carried out in which the explicit form of the electron spectrum in GSL was received.
This result was used in \cite{33,34} to study the influence of THz radiation on the current-voltage characteristic of GSL.

Below the electron energy in GSL is obtained in explicit form by using of tight binding model.
Using this dispersion low the EM wave equation is shown to be a solution of sine-Gordon (SG) equation in collisionless approximation in GSL.
Last equation has the Jacobi elliptic functions as a general periodic solution.
To study the form of nonlinear EM wave in GSL the effect of mutual rectification of waves can be used.
Therefore the transversal direct current induced by the cnoidal and sinusoidal waves with orthogonal polarization lines under the longitudinal constant electric field is investigated below.
Direct current induced by mixing of these waves is shown to change its direction when the longitudinal electric field changes its absolute value.
\\

\begin{center}\textbf{II. Electron spectrum of GSL}\\\end{center}

\begin{figure}[t]
\centering
\includegraphics[width=1\linewidth]{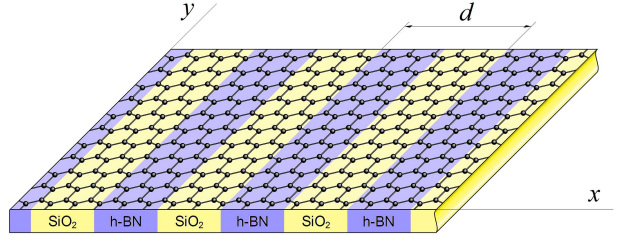}
\caption{\textbf{Fig. 1.} GSL scheme}
\label{fig1}
\end{figure}

\begin{figure}[b]
\centering
\includegraphics[width=1\linewidth]{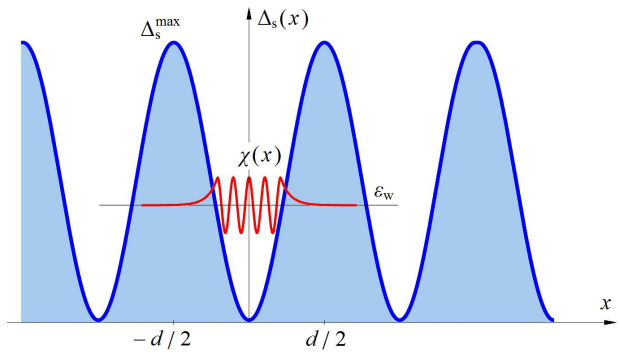}
\caption{\textbf{Fig. 2.} Scheme of a periodic modulation of the band gap}
\label{fig2}
\end{figure}

We consider a superlattice obtained by a sheet of graphene deposited on a banded substrate formed by periodically alternating layers of SiO$_{2}$ and h-BN as it is shown in \hyperref[fig1]{Fig. 1}.
The layers are arranged so that the hexagonal crystal lattice of h-BN was under the hexagonal lattice of graphene.
Due to this, in the areas of graphene plane located above the layers of h-BN an energy gap is equal to 0.053 eV arises.
The result is a periodic modulation of the band gap \cite{9}.
If the layers of SiC are used then the value of the gap is 0.26 eV.
\end{multicols}
\twocolumn
The band gap width is proposed to have the arbitrary periodic profile $\Ds{\left( x\right) }=\Ds{\left( x+d\right) }$ (\hyperref[fig2]{Fig. 2}) unlike that of \cite{9}.
Here $d$ is the SL period.
Electron states are described by the spinor function
$$
\psi=\begin{pmatrix} \mspace{5mu} \psi_\uparrow\left(\mathbf{p}, \mathbf{r}\right)  \mspace{5mu} \\
\mspace{5mu} \psi_\downarrow\left(\mathbf{p}, \mathbf{r}\right) \mspace{5mu} \end{pmatrix} \com
$$
where $\psi_\uparrow\left(\mathbf{p}, \mathbf{r}\right)$ and $\psi_\downarrow\left(\mathbf{p}, \mathbf{r}\right)$ describe the electron states in the first and second crystal sublattices (called also as pseudospins), respectively \cite{1,2}.
Dirac equation for the function $\psi\left(\mathbf{p}, \mathbf{r}\right)$ has the form as follows \cite{9}
\begin{gather}\label{1}
\vF \hat {\boldsymbol{\sigma}}\cdot \hat {\mathbf{p}} \psi+\Ds{\left( x\right) }\s{z}\psi=\varepsilon \psi \pnt
\end{gather}
Here $\hat{\boldsymbol{\sigma}}=\left( \s{x}\com \mspace{15mu}\s{y}\right)$, $\s{z}$ are the Pauli matrixes, $\hat{\mathbf{p}}=-i\boldsymbol{\nabla}$, $\vF$ is the velocity on the Fermi surface.
Acting twice on the spinor $\psi$ by the Hamiltonian
$\hat {\mathscr{H}}=\vF \hat {\boldsymbol{\sigma}}\cdot \hat {\mathbf{p}} +\Ds{}\s{z}$
we derive instead of (\ref{1})
\begin{gather}\label{2}
\vF^2\left( \p{x}^2+\p{y}^2 \right)\psi -\vF\dfrac{d\Ds{}}{dx}\s{y}\psi+\Delta_\text{s}^2\left( x\right)\psi=\varepsilon^2\psi \pnt
\end{gather}
The equation (\ref{2}) admits the solutions which can be written in the form
$\psi\left( \mathbf{r} \right)= F\left( x\right)\exp\left( ip_yy\right) $,
where $F$ is the column-spinor
$$
F=\begin{pmatrix} \mspace{5mu} F_\uparrow\left(x\right)  \mspace{5mu} \\
\mspace{5mu} F_\downarrow\left(x\right) \mspace{5mu} \end{pmatrix} \pnt
$$
which satisfies the Bloch theorem.
So we can write
$F\left( x+d \right)= F\left( x\right)\exp\left( ip_xd\right)  $.
The next equation is obtained for the spinor $F$ 
\begin{gather}\label{3}
-\vF^2\dfrac{d^2F}{dx^2} -\vF\dfrac{d\Ds{}}{dx}\s{y}F+\Delta_\text{s}^2 F=
\left( \varepsilon^2 - \vF^2 p_y^2\right) F \pnt
\end{gather}
The probability of electron penetration through the barrier between the wells is suggested to be small.
In this case spinor $F\left( x\right) $ can be represented as the following linear combination
\begin{gather}\label{4}
F\left( x\right) =\sum\limits_n\chi\left( x-nd\right) e^{inp_xd} \com
\end{gather}
where $\chi\left( x\right) $ is the column-spinor
$$
\chi=\begin{pmatrix} \mspace{5mu} \chi_\uparrow\left(x\right)  \mspace{5mu} \\
\mspace{5mu} \chi_\downarrow\left(x\right) \mspace{5mu} \end{pmatrix} \pnt
$$
which describes the one-dimensional motion of electron with the energy inside one of the wells  $\ew$.
Its components decay exponentially on either side of the wells borders.
This spinor obeys the equation
\begin{gather}\label{5}
\vF \s{x}\p{x} \chi\left( x\right) +\Ds{\left( x\right) }\s{z}\chi\left( x\right) =\ew \chi\left( x\right)  \pnt
\end{gather}
After some transformations which are similar above we obtain instead of (\ref{5})
\begin{gather}\label{6}
-\vF^2\dfrac{d^2\chi}{dx^2} -\vF\dfrac{d\Ds{}}{dx}\s{y}\chi+\Delta_\text{s}^2 \chi=\ew^2 \chi \pnt
\end{gather}

Now we multiply the equation (\ref{3}) on the left by the row-spinor
$\chi^+=\left( \chi_\uparrow^* \com \mspace{15mu} \chi_\downarrow^*\right) $
and the equation (\ref{5}) on the left by the row-spinor
$F^+=\left( F_\uparrow^* \com \mspace{15mu} F_\downarrow^*\right) $.
As a result we obtain
\begin{multline}\label{7}
-\vF^2\chi^+\dfrac{d^2F}{dx^2} -\vF\dfrac{d\Ds{}}{dx}\chi^+\s{y}F+\Delta_\text{s}^2\chi^+ F \\
=\left( \varepsilon^2 - \vF^2 p_y^2\right) \chi^+F \com
\end{multline}
\begin{multline}\label{8}
-\vF^2F^+\dfrac{d^2\chi}{dx^2} -\vF\dfrac{d\Ds{}}{dx}F^+\s{y}\chi \\
+\Delta_\text{s}^2 F^+\chi=\ew^2 F^+\chi \pnt
\end{multline}
After taking the Hermitian conjugation in the both sides of (\ref{8}) and taking into account the next relations
$$
\left(F^+\chi \right)^+=\chi^+F \com \mspace{15mu} \left(F^+\s{y}\chi \right)^+=\chi^+\s{y}F
$$
we rewrite (\ref{8})
\begin{multline}\label{9}
-\vF^2\dfrac{d^2\chi^+}{dx^2}F -\vF\dfrac{d\Ds{}}{dx}\chi^+\s{y}F \\
+\Delta_\text{s}^2 \chi^+F=\ew^2 \chi^+F \pnt
\end{multline}
Subtracting equations (\ref{7}) and (\ref{9}) we arrive at
\begin{multline}\label{10}
-\vF^2\dfrac{d}{dx}\left(\chi^+\dfrac{dF}{dx}-\dfrac{d\chi^+}{dx}F \right) \\
=\left( \varepsilon^2-\ew^2-\vF^2p_y^2\right) \chi^+F \pnt
\end{multline}
\begin{figure*}[t]
\centering
\includegraphics[width=0.9\linewidth]{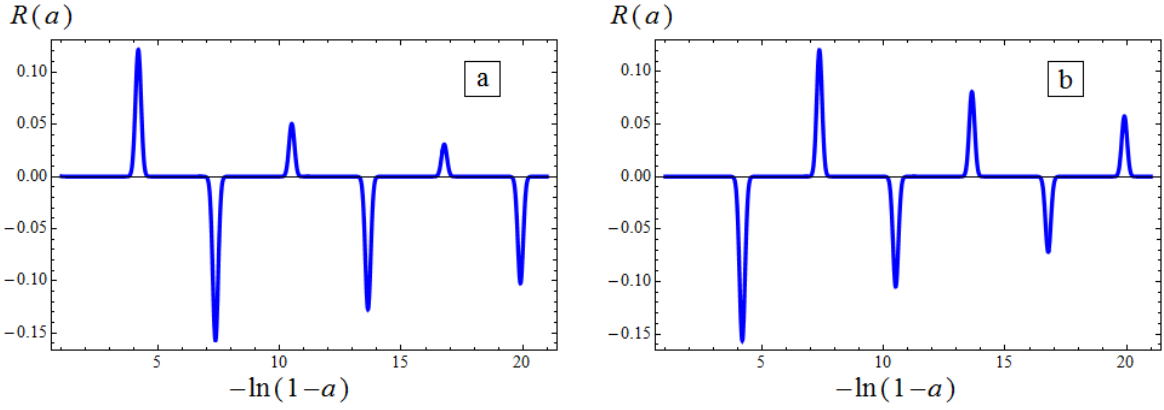}
\caption{\textbf{Fig. 3.} Function $R(a)$ vs the cnoidal wave amplitude (logarithmic scale). $a<1$, $\Omega_0\tau=10, \omega\tau=10$, (a) $\fB{}\tau=1$, (b) $\fB{}\tau=11$}
\label{fig3}
\end{figure*}

After integration of (\ref{10}) over the SL period and using the combination (\ref{4}) we derive\\
\begin{multline}\label{11}
-\vF^2\sum\limits_n e^{inp_xd}\left(\Large\mathstrut \chi^+\left( x\right) \varphi\left( x-nd\right)\right. \\ 
\mspace{150mu}\left.\left.\Large\mathstrut-\varphi^+\left( x\right)\chi\left( x-nd\right)  \right) \right|_{-d/2}^{+d/2} \\
=\left( \varepsilon^2-\ew^2-\vF^2p_y^2\right)\sum\limits_n e^{inp_xd} \mspace{150mu} \\
\times\int\limits_{-d/2}^{+d/2}\chi^+\left( x\right) \chi\left( x-nd\right) dx \com 
\end{multline}
where we define $\varphi=\partial_x\chi$.
Due to the weak transparency of the SL barriers we have
$$
\left| \int\limits_{-d/2}^{+d/2}\chi^+\left( x\right) \chi\left( x-nd\right) dx\right| \ll 1 \com
\mspace{15mu} \text{when } n\neq0 \pnt
$$
Besides we take into account the next circumstances.
Firstly, if $\ew$ corresponds to the lowest energy level of electron in the well then components of $\chi$ are even functions.
Hence their derivations are odd functions.
Secondly, due to the strong damping of the component of $\chi$ as electron moves into neighbor wells the next inequalities are performed
$$
\left| \Large\mathstrut \chi^+\left( nd\right) \chi\left( kd\right) \right| \ll
\left| \Large\mathstrut \chi^+\left( 0\right) \chi\left( 0\right) \right| \com \mspace{15mu}
\left| \Large\mathstrut \chi^+\left( 0\right) \chi\left( \pm d\right) \right| \com
$$
$$
\text{when } k\com \mspace{15mu} n\neq0 \pnt
$$
The latter allows to leave in the left-side sum of (\ref{11}) the terms with $n=$0, 1 for $x=d/2$ and the terms with $n=$0, $-$1 for $x=-d/2$.
As a result we obtain
\begin{gather}\label{12}
\varepsilon^2-\ew^2-\vF^2p_y^2
=4\vF^2\chi^+\left( \dfrac{d}{2}\right) \varphi\left(\dfrac{d}{2}\right)\cos p_xd \com 
\end{gather}

For the components of the spinor $\chi$ the quasiclassical expressions of the wave functions corresponding to the motion under the barrier ($x>0$) should be chosen.
They can be rated approximately as
$$
\chi\sim\dfrac{1}{\sqrt{d}}e^{-\kappa_0x} \com \mspace{15mu}
\varphi\sim-\dfrac{\kappa_0}{\sqrt{d}}e^{-\kappa_0x} \com
$$
where $\kappa_0\sim\Delta_\text{s}^\text{max}/\vF$, $\Delta_\text{s}^\text{max}$ is the barrier height.

Finally we rewrite (\ref{12}) in the form
\begin{gather}\label{13}
\varepsilon\left(\mathbf{p} \right) =\pm\sqrt{\Gamma^2+\vF^2p_y^2
-\gamma^2\cos p_xd} \com
\end{gather}
where parameters  $\Gamma$   and  $\gamma$   are defined by the shape of SL barriers and by the ratio between the barrier width and the well width.
Further the valence miniband is proposed to be completely filled by the electrons so the conduction miniband is considered only.

The low transparency of barriers ($\gamma \ll \Gamma$) allows the expression (12) to be written in the next approximate form
\begin{gather}\label{14}
\varepsilon\left(\mathbf{p} \right) =\sqrt{\Gamma^2+\vF^2p_y^2}
	-\dfrac{\gamma^2\cos p_xd}{2\sqrt{\Gamma^2+\vF^2p_y^2}} \com
\end{gather}

In \cite{33} the following explicit expression for the dimensionless energy of electron in GSL was matched by the numerical analysis of the dispersion low \cite{9} written in the case when widths of barrier and well are equal.
\begin{multline}\label{15}
\varepsilon_\text{numer}=\sqrt{1+\left(f_3/f_2 \right)^2q_y^2} \\
+\dfrac{f_4}{f_2^2}\dfrac{1-\cos q_x}{\sqrt{1+\left(f_3/f_2 \right)^2q_y^2}}  \com
\end{multline}
where $\mathbf{q}=\mathbf{p}d$, $f_3/f_2\sim$0.58, $f_4/f_2^2\sim$0.24.
Let expression (\ref{13}) is transformed to dimensionless form analogous to (\ref{15})
\begin{gather}\label{16}
\varepsilon_\text{theor}=\sqrt{1+\alpha_1^2q_y^2}
+\dfrac{\alpha_2\left( 1-\cos q_x\right) }{\sqrt{1+\alpha_1^2q_y^2}}  \com
\end{gather}
For the square profile of the SL potential with equal widths of the barrier and well and for the barrier height of 0.027 eV corresponding to the substrate h-BN/SiO$_{2}$ the values of  $\alpha_1$ and   $\alpha_2$ are 0.69 and 0.19 correspondingly.
It coincides with the result \cite{33} nearly.

It should be noted, that the approximation of weak barriers transparency used to obtain the formula (\ref{14}) the better, the more the gap arising in GSL. Therefore, the formula (\ref{14}) is more valid if SiC is chosen instead h-BN for the periodic substrate. In addition, as pointed out in \cite{33}, this choice is more suitable for a single-mini-band applications.
\begin{figure}[t]
\centering
\includegraphics[width=1\linewidth]{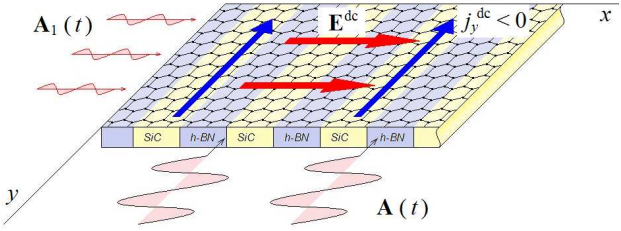}
\caption{\textbf{Fig. 4.} Schematic of the process. Transversal current is negative}
\label{fig4}
\end{figure}
\\

\begin{center}\textbf{III. EM waves in the ideal GSL}\\\end{center}

Let an EM wave propagate along the graphene plane so that the vector of the electric field intensity is directed along the $Oz$.
In the collisionless approximation, the current density is calculated with the next formula
\begin{gather}\label{17}
j_x=-e\sum\limits_\mathbf{p}v_x\left( \Large\mathstrut \mathbf{p}+e\mathbf{A}\left( y,t\right) \right) f_0\left( \mathbf{p}\right) \com
\end{gather}
where $\mathbf{A}\left( y,t\right) $ is the vector potential of the EM wave field, $f_0\left( \mathbf{p}\right) $ is the equilibrium state function, and $\mathbf{v}=\partial_\mathbf{p}\varepsilon$ is the electron velocity along the GSL axis. After calculating the velocity $v_z$ with the help of (\ref{14}), we obtain
\begin{gather}\label{18}
j_x\left(t \right) =-\dfrac{e\vF\gamma^2}{2\Gamma}\sum\limits_\mathbf{p}f_0\left( \mathbf{p}\right)
\dfrac{\sin \left(q_x+\mathcal{A} \right)}{\sqrt{1+\beta^2q_y^2}} \com
\end{gather}
where $\mathcal{A}=edA_x$, and the parameter $\beta=\vF/\Gamma d$ has the order of 1.
After calculation of the sums by momentums at low temperatures ($\theta<<\Gamma$), we have
\begin{gather}\label{19}
j_x\left(t \right) =-\dfrac{e n_0 \vF\gamma^2}{2\Gamma}\sin \mathcal A \pnt
\end{gather}
Here $n_0$ is the surface concentration of the charge carriers.
Using formula (\ref{19}) we write the d’Alembert equation
\begin{gather}\label{20}
\dfrac{\partial^2\mathcal A}{\partial t^2} - \dfrac{\partial^2\mathcal A}{\partial y^2} +\fpl{2}\sin \mathcal A=0 \com
\end{gather}
where $\fpl{}$ is plasma frequency
$$
\fpl{}=ed\gamma\sqrt{\dfrac{2\pi n_0}{\Gamma}} \pnt
$$
Jacobi elliptic functions are the general periodic solution of equation (\ref{20}).
Here the wavelength is assumed to be much more than the electron length of free path. Hence the coordinate part of the function
$\mathcal A\left(y,t \right) $ is neglected.
The solution of (\ref{20}) can be written by using the Fourier series
\begin{gather}\label{21}
e^{i\mathcal A\left( t\right) }=\sum\limits_{n=-\infty}^{\infty}c_n\left( a\right) e^{in\Omega\left( a\right) t} \pnt
\end{gather}
Here in the case $0<a<1$ the next relations should be used
$$
\Omega\left( a\right) =\dfrac{\pi\Omega_0}{2K\left( a\right) } \com \mspace{15mu}
c_0=\dfrac{2E\left( a\right) }{K\left( a\right)}-1 \com
$$
$$
c_n=\dfrac{n\pi^2}{K^2\left( a\right)}\dfrac{1}{q_a^{-n/2}+\left(-1 \right)^{n+1}q_a^{n/2}} \com
$$
and in the case $a>1$ one should use
$$
\Omega\left( a\right) =\dfrac{\pi\Omega_0a}{2K\left( a^{-1}\right) } \com \mspace{15mu}
c_0=1-2a^2+\dfrac{2a^2E\left( a^{-1}\right) }{K\left( a^{-1}\right)} \com
$$
$$
c_n=\dfrac{4n\pi^2a^2}{K^2\left( a^{-1}\right)}\dfrac{1}{q_{1/a}^{-n}-q_{1/a}^{3n}} \pnt
$$
In both cases $\Omega_0=\fpl{}u\sqrt{\left|1-u^2 \right| } \com$ $a=eE_0d/2\Omega_0$, $E_0$ is the amplitude of the EM wave described by the potential $\mathcal A$,
$$
q_\xi=\exp\left(-\dfrac{\pi K\left(\sqrt{1-\xi^2} \right) }{K\left( \xi\right) } \right) \com
$$ 
$u$ is the wave velocity, $K(\xi)$ and $E(\xi)$ are complete elliptic integrals of the first and second kind correspondingly.
\begin{figure}[t]
\centering
\includegraphics[width=1\linewidth]{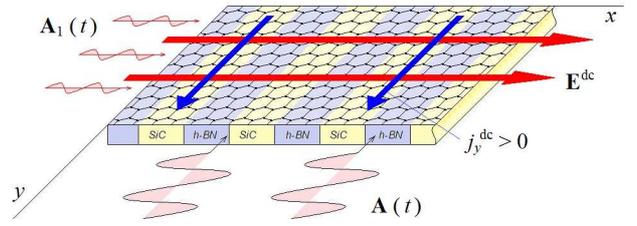}
\caption{\textbf{Fig. 5.} Schematic of the process. Transversal current is positive}
\label{fig5}
\end{figure}
\\

\begin{center}\textbf{IV. The mutual rectification of the sinusoidal and cnoidal EM waves in GSL}\\\end{center}
\begin{figure*}[t]
\centering
\includegraphics[width=0.9\linewidth]{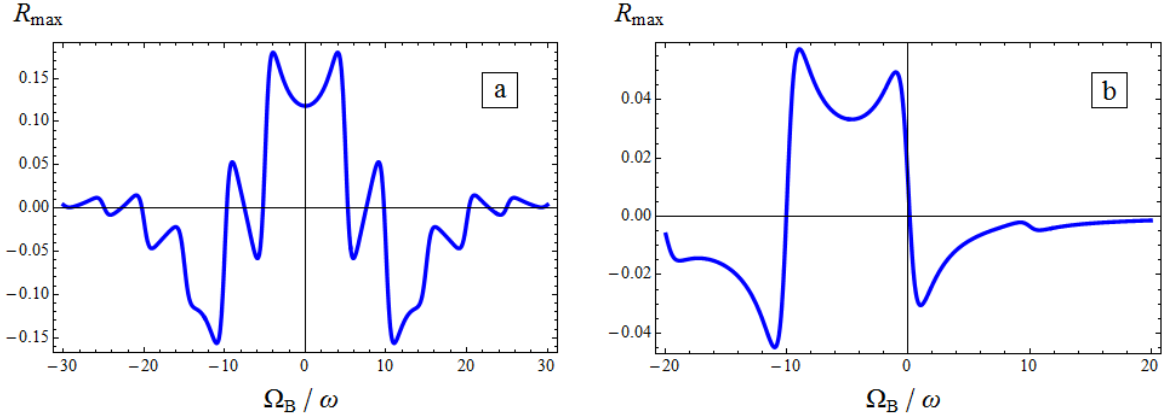}
\caption{\textbf{Fig. 6.} Transversal direct current vs longitudinal electric field intensity. $\Omega_0\tau=10, \omega\tau=10$, (a) $a=0.984$, (b) $a=1.014$}
\label{fig6}
\end{figure*}

The direct current arising in GSL due to wave mixing was studied in \cite{33,35}.
The influence of longitudinal electric field on the transversal direct current induced in GSL by two EM waves was investigated in \cite{33} where the wave with electric field intensity applied along the SL axis was supposed to be of sinusoidal form as well as in \cite{31,32} but not of cnoidal form.
The mutual rectification of sinusoidal and cnoidal  waves with orthogonal polarization lines in GSL was investigated in \cite{35} where the direct current induced by these waves through to the GSL axis in absence of the constant electric field was calculated.

Here we calculate the direct current induced by sinusoidal and cnoidal EM waves with orthogonal polarization lines perpendicularly to the GSL axis in presence of the constant electric field.
Direction of this current is shown below to be regulated by changing of absolute value of the electric field intensity.

The constant electric field with intensity is equal to
$\mathbf E^{\text{dc}}=\left(E_x^{\text{dc}} \com \mspace{15mu} 0 \com \mspace{15mu} 0\right)$
is suggested to be applied along the SL axis.
The EM wave with vector potential
$\mathbf A=\left(A_x \com \mspace{15mu} 0 \com \mspace{15mu} 0\right)$
is supposed to be polarized along the GSL axis.
Hence it has the cnoidal form satisfying to the SG equation.
Another wave with vector potential
$\mathbf A_1=\left(0 \com \mspace{15mu} A_y \com \mspace{15mu} 0\right)$
is supposed to be weak and to be polarized perpendicularly to the GSL axis.
It has the sinusoidal form
\begin{gather}\label{22}
A_{1y}=-\dfrac{E_{01}}{\omega}\sin\left( \omega t+\varphi\right) \com
\end{gather}
where $E_{01}$, $\omega$, $\varphi$ are amplitude, frequency and initial phase of the transversal field oscillations.
Under these conditions and in the constant relaxation time $\tau$ approximation the electric current density arising through the axis $Oy$ is calculated with the following formula
\begin{multline}\label{23}
j_y\left( t\right) =-e\int\limits_0^\infty d\xi \mspace{3mu} e^{-\xi}\sum\limits_\mathbf{p}f_0\left(\mathbf{p}\right)  \\
\times v_y\left(\Large\mathstrut \mathbf{p}+e\left(\mathbf A-\mathbf A'
+\mathbf A_1-\mathbf A'_1 \right)-e\mathbf E\tau\xi \right)  \com
\end{multline}
where
$\mathbf A'=\mathbf{A}\left(t-\tau\xi\right) \com \mspace{15mu}$
$\mathbf A'_1=\mathbf{A_1}\left(t-\tau\xi\right) \pnt$

The formula (\ref{23}) is rewritten in the next form
\begin{multline}\label{24}
j_y\left( t\right) =-e\vF\int\limits_0^\infty d\xi \mspace{3mu} e^{-\xi}\sum\limits_\mathbf{p}f_0\left(\mathbf{p}\right) \\
\times\dfrac{q_y+\mathcal B-\mathcal B'}{\sqrt{1+\left( q_y+\mathcal B-\mathcal B'\right) ^2}}   \\
\times\left(1+\dfrac{\gamma^2}{2\Gamma^2}\dfrac{\cos \left(q_x-\fB{}\tau\xi+\mathcal A-\mathcal A'\right) }{1+\left( q_y+\mathcal B-\mathcal B'\right) ^2} \right)\pnt
\end{multline}
Here $\mathcal B=edA_{1y} \com$ $\mathcal B'=edA'_{1y} \com$
$\fB{}=eE_x^{\text{dc}}d$ is the Bloch oscillations frequency.
Further the sinusoidal EM wave is supposed to be weak ($b \equiv e\vF E_{01}/\omega\Gamma\ll 1$).
In the linear approximation in the parameter $b$ and at low temperatures ($\theta \ll \Gamma$) we obtain the following formula for the transverse direct current density
\begin{gather}\label{25}
j_y^{\text{dc}}=-\dfrac{\gamma^2}{2\Gamma^2\omega\tau} \sigma_\bot E_{01} R\left( a\right) \cos\varphi \com
\end{gather}
where $\sigma_\bot=e^2 n_0 \tau/m_\bot \com \mspace{15mu}$ $m_\bot=\Gamma/\vF^2 \com$
\begin{multline}\label{26}
R\left( a\right)=\sum\limits_{k,n=-\infty}^\infty c_kc_n\left(\delta_{n\Omega,k\Omega-\omega}
-\delta_{n\Omega,k\Omega+\omega} \right) \\
\times\dfrac{\left(\fB{}+n\Omega \right)\tau}{1+\left( \fB{}+n\Omega\right)^2 \tau^2} \pnt
\end{multline}
and $\delta_{n,k}$ is Kronecker delta.
\\

\begin{center}\textbf{V. Discussion}\\\end{center}

The graphics of the dependence of the function (26) on the cnoidal wave amplitude a under the different values of the longitudinal electric field intensity is shown in \hyperref[fig3]{Fig. 3} ($0<a<1$).
This function can be seen to have the resonant character.
The constant component of the transversal current occurs whenever the sinusoidal wave frequency is the integer number of cnoidal wave frequency.
The possibility of the direct current in the transverse direction with relation to the SL axis is explained by the nonadditivity of the GSL spectrum.
In the bulk SL with additive spectrum such effect is impossible.

In the absence of the longitudinal electric field ($E_x=0$) direct current resonances arise when $\omega$   is in even number more than $\Omega$.
However when longitudinal electric field is present ($E_x\neq0$) the new resonances of transversal direct current appears.
Moreover, the resonances values and resonances directions can be regulated by changing of absolute value of the longitudinal electric field (Figs. \ref{fig4}, \ref{fig5}).
This effect is also possible due to nonadditivity of the electron spectrum (\ref{14}).
The direct current dependence on the longitudinal electric field intensity is shown in \hyperref[fig6]{Fig. 6}.

In the case $a>1$ the function $R_\text{max}\left( \fB{}\right) $ can be seen from \hyperref[fig6]{Fig. 6} (b) to be not symmetric, unlike the case of $0<a<1$ (\hyperref[fig6]{Fig. 6}, a).
This fact is due to the constant component of electric field of cnoidal wave in the case when $a>1$.
\\

\begin{center}\textbf{Acknowledgements}\\\end{center}

This study was supported by the Administration of the Volgograd Region (State Research Project).\\
\onecolumn
\begin{multicols}{2}

\end{multicols}
\end{document}